\documentclass[preprint,12pt]{elsarticle}

\usepackage{amssymb}
\usepackage{subfigure}
\usepackage[mathlines]{lineno}

\def\ifb{\mbox{fb$^{-1}$}}
\def\cm{\ifmmode  {\mathrm{cm}}\else
                   \textrm{cm}\fi}%
\def\TeV{\ifmmode {\mathrm{\ Te\kern -0.1em V}}\else
                   \textrm{Te\kern -0.1em V}\fi}%
\def\GeV{\ifmmode {\mathrm{\ Ge\kern -0.1em V}}\else
                   \textrm{Ge\kern -0.1em V}\fi}%
\def\MeV{\ifmmode {\mathrm{\ Me\kern -0.1em V}}\else
                   \textrm{Me\kern -0.1em V}\fi}%
\def\keV{\ifmmode {\mathrm{\ ke\kern -0.1em V}}\else
                   \textrm{ke\kern -0.1em V}\fi}%
\def\eV{\ifmmode  {\mathrm{\ e\kern -0.1em V}}\else
                   \textrm{e\kern -0.1em V}\fi}%

\begin{document}

\begin{frontmatter}



\title{Construction of the new silicon microstrips tracker for the Phase-II ATLAS detector}

\author[label0]{Zhijun Liang\footnote{ email: liangzj@ihep.ac.cn} } 
\author[]{for the ATLAS collaboration} 

\address[label0]{ State Key Laboratory of Particle Detection and Electronics (Institute of High Energy Physics, CAS)，Beijing 100049 }
  
%
%
%
\begin{abstract}

  

  In next ten years, the Large Hadron Collider will be upgraded to the High Luminosity LHC (HL-LHC), resulting in ten time more integrated luminosity.  To withstand the much harsher radiation and occupancy conditions of the HL-LHC, the inner tracker of the ATLAS detector must be redesigned and rebuilt completely. 
    The design of the ATLAS Upgrade inner tracker (ITk) has already been defined. It consists of several layers of silicon particle detectors. The innermost layers will be composed of silicon pixel sensors, and the outer layers will consist of silicon microstrip sensors. This paper will focus on the latest research and development activities performed by ITk strips community with respect to the assembly and test of the strip modules and the stave and petal structures. 

\end{abstract}


\end{frontmatter}


\section{Introduction to ATLAS phase-II upgrade}
The inner detector of the present ATLAS detector has been designed and developed to function in the environment of the present Large Hadron Collider (LHC)~\cite{ID}. The next major upgrade phase of the Large Hadron Collider (LHC) is  currently foreseen to be completed in 2026~\cite{HLLHC}. It is called the High Luminosity-LHC (HL-LHC), and it aims to increase the integrated luminosity to about ten times the original LHC design luminosity, resulting in an additional integrated luminosity of 3000~\ifb -- 4000~\ifb over ten years. These data will improve the precision of the measurement of the Higgs properties and enhance the sensitivity to search for new physics. 

Extracting meaningful physics in HL-LHC collisions environment requires a new inner detector. The new detectors must be faster, they need to be more highly segmented, and covering more area. They also need to be more resistant to radiation, and they require much greater power delivery to the front-end systems.  For those reasons, the inner tracker of the ATLAS detector must be redesigned and rebuilt completely. The design of the ATLAS Upgrade inner tracker (ITk) has already been defined in ATLAS strip detector upgrade technical design report~\cite{TDR}. The new ATLAS inner tracker will be an all-silicon tracker. As shown in Fig.~\ref{fig:layout}, The ITk strip detector consists of 4 barrel layers extending to $z =\pm$1400~$\mathrm{mm}$ and 12 endcap disks (6 on each side) extending from the edge of the barrel to  $z =\pm$3000~$\mathrm{mm}$. The innermost barrel layers will be populated with short strip segments (24.1~$\mathrm{mm}$) to increase granularity at small radius as compared to the long strip segments (48.2~$\mathrm{mm}$) in the outer barrel. The total area of ITK strip detector is 165~$\mathrm{m}^2$.


The biggest change to the current ATLAS inner tracker is the replacement of the Transition Radiation Tracker(TRT) with 47.8 $\mathrm{mm}$ long silicon strips.
The outer active radius is slightly larger, improving momentum resolution. The number of strip modules and readout channels will be increased by one order of magnitude compared to current ATLAS strip detector in this baseline design.


\begin{figure}[]
  \centering
  \includegraphics[width=0.6\textwidth]{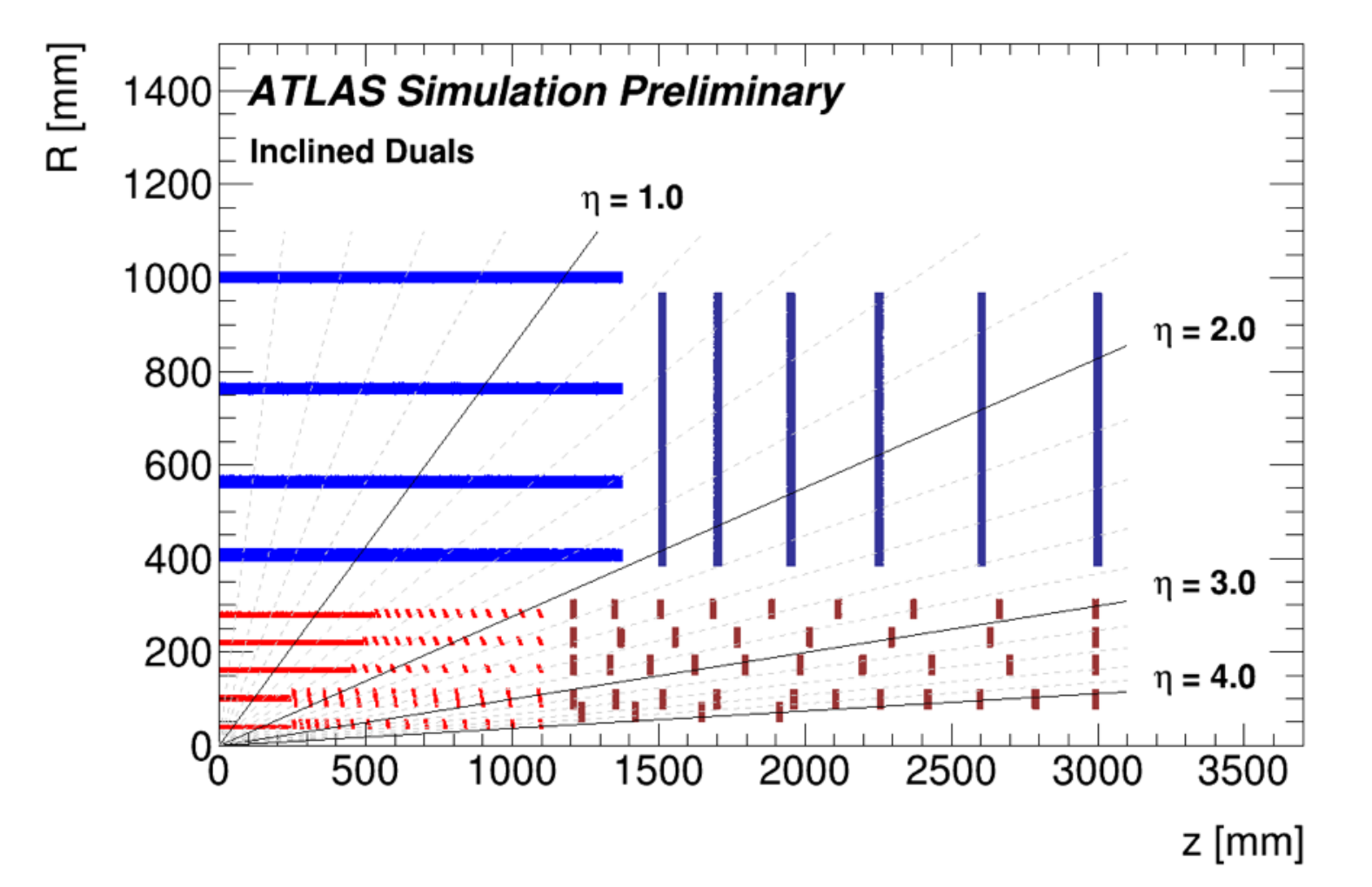}
  
  \caption{ The layout of the ITk detector. Here only one quarter segment is shown. The horizontal axis, $z$, is the axis along the beam line with zero being the interaction point. The vertical axis, $R$, extends outward perpendicular to the beam pipe and the origin of co-ordinates is taken to be the center of the detector~\cite{TDR}.}
\label{fig:layout}
\end{figure}

\section{Staves and Petals: The basic building block}

In response to the needs of the strip region for the ITk, highly modular structures are being studied and developed, called "staves" for the central region (barrel) and "petals" for the forward regions (end-caps). These structures are designed to minimise material for large scale assembly and easier replacement~\cite{TDR,staveref} and they integrate large numbers of sensors and readout electronics, with precision light weight mechanical elements and cooling structures.

Strip modules are mounted on either side of the stave (petal)  with the strips on opposite sides at a small stereo angle to provide space points to be used for tracking. Staves (petals) are populated with 14 (9) strip modules on each side. 
Each module sensor is glued to a bus tape for power and communications. Electrical power and signals are received by the stave (petal) at a board called the End Of Substructure Cards (EoS). The bus tape is co-cured to the carbon fibre frame, which provides mechanical stability.The carbon fibre frame is called the stave/petal core, which is a core of carbon fibre honeycomb and carbon foam with embedded cooling pipes sandwiched between two carbon fibre facings, as shown in Fig.~\ref{fig:stavecore}. 

\section{Baseline layout for new ATLAS tracker design }

\begin{figure}[]
  \centering
  \includegraphics[width=0.7\textwidth]{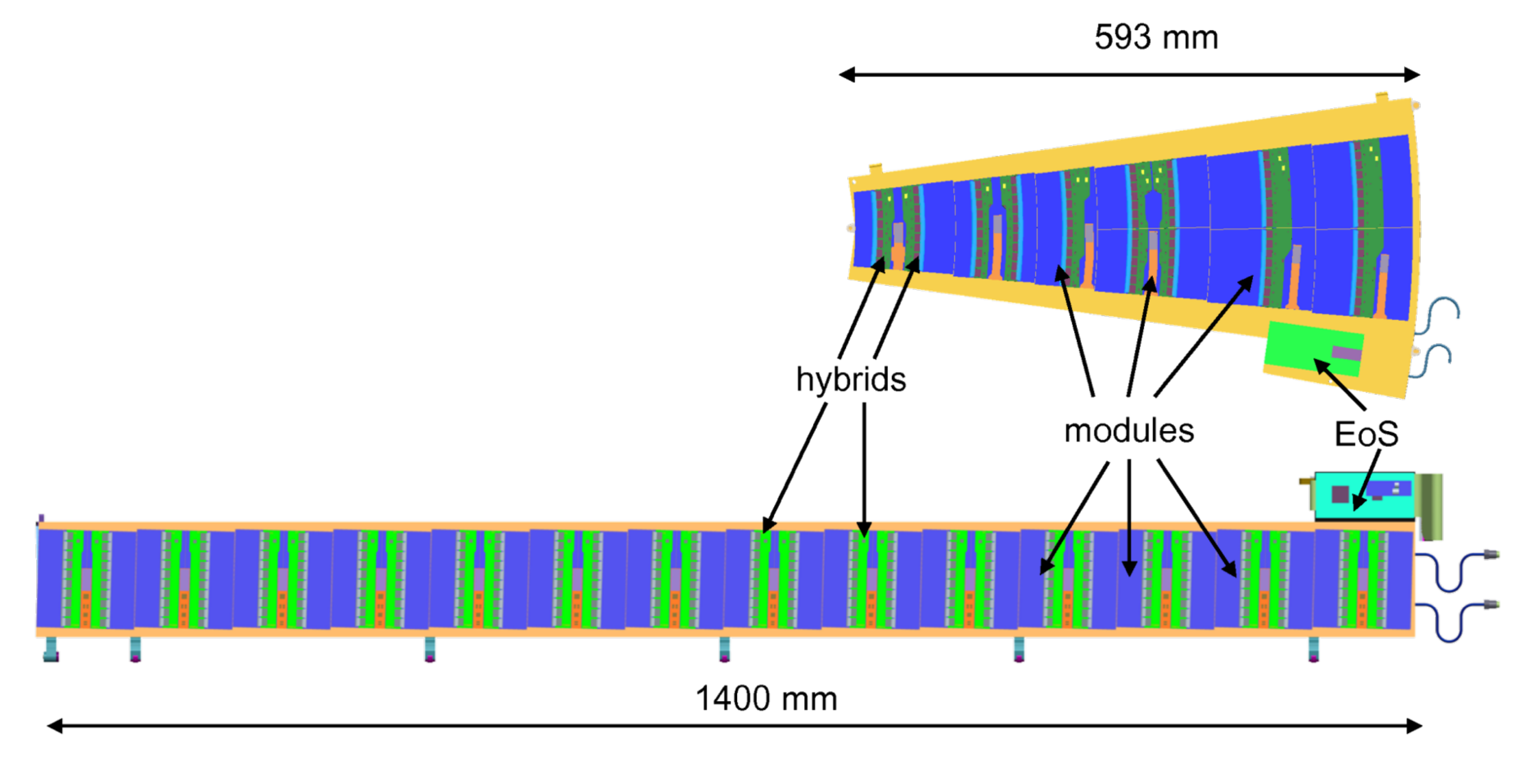}
  
  \caption{Petal (top) and stave (bottom) structures showing the placement of modules, the End of Substruc- ture Cards (EoS) and the cooling pipe connections.
}
\label{fig:stavepetal}
\end{figure}

Staves are arranged in concentric cylinders centred on the beam-line as illustrated in Fig.~\ref{fig:stavelayout} (a). The staves are rotated by a small angle (the tilt-angle) away from the tangent to allow an overlap in the $\phi$ direction. 
The arrangement for the endcap modules on the supporting structure is indicated in Fig.~\ref{fig:stavelayout} (b). There are six disks on each forward side with 32 petals per disk.

\begin{figure}[]

  \centering
    \includegraphics[width=0.65\textwidth]{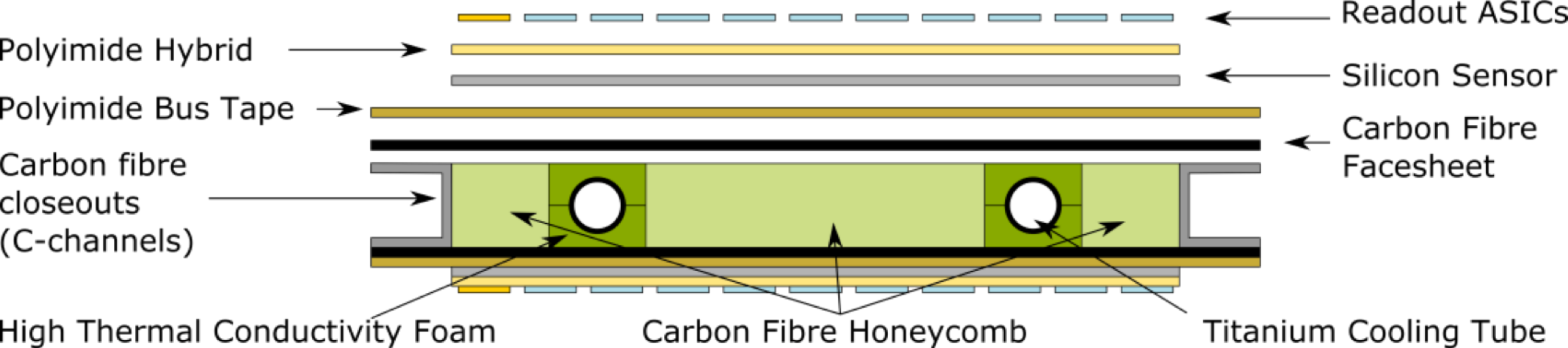}
  \caption{Schematic (not to scale) showing the internal structure of a stave core~\cite{TDR}.}
\label{fig:stavecore}
\end{figure}

\begin{figure}[]

  \centering
    \subfigure[]{\includegraphics[width=0.27\textwidth]{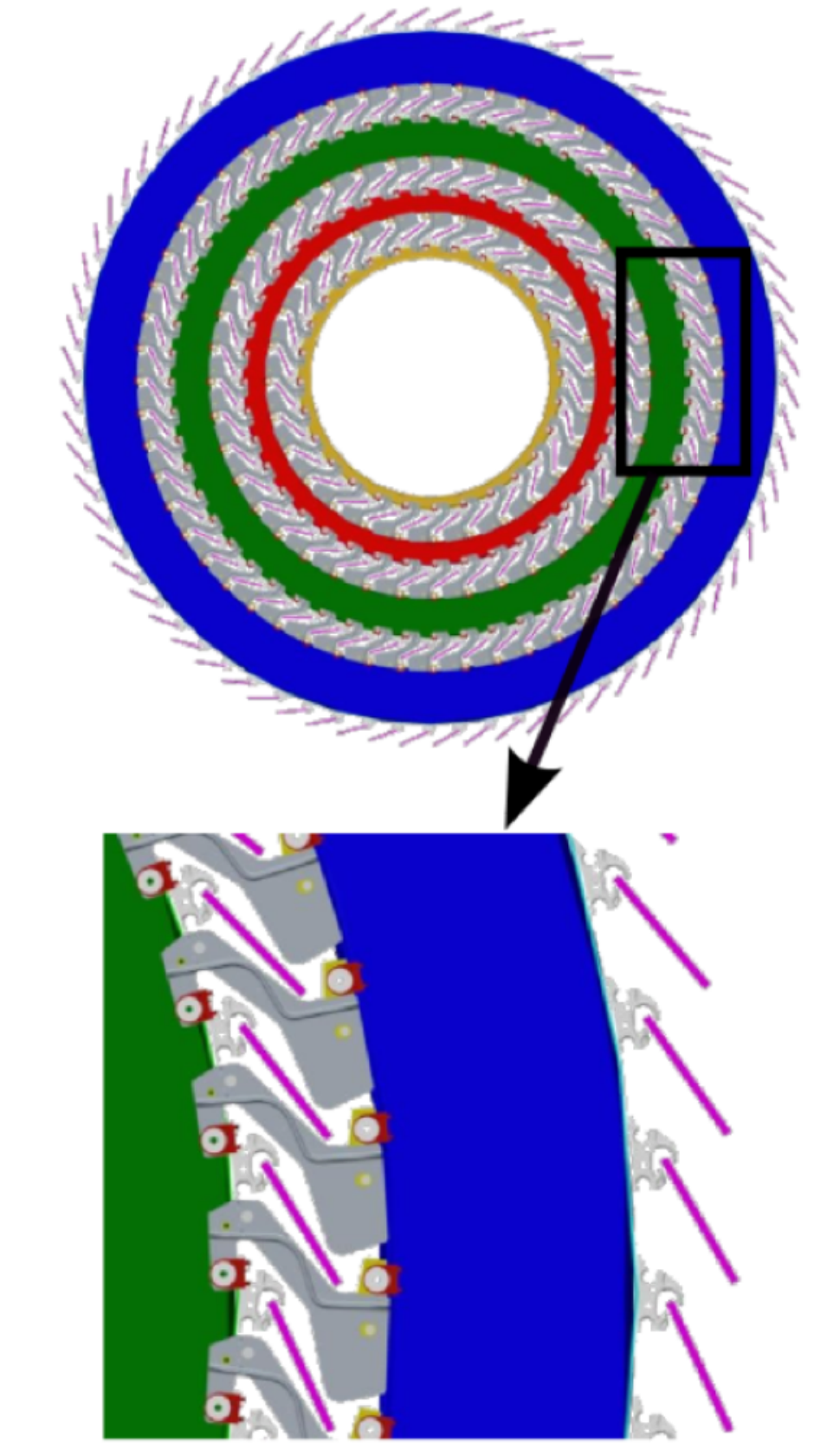}}
   \subfigure[]{\includegraphics[width=0.49\textwidth]{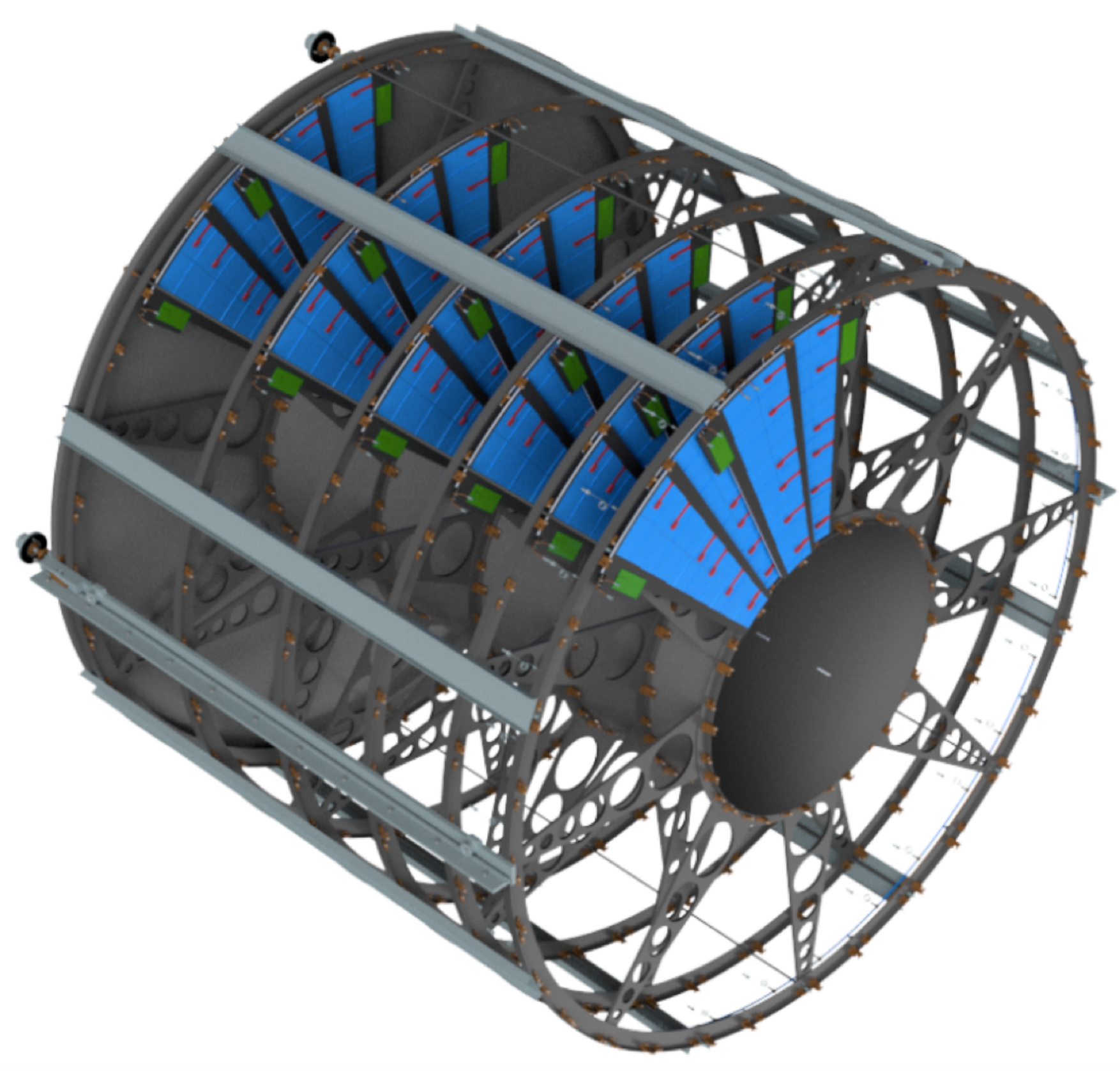}}
  \caption{(a) Arrangement of staves in barrels.  (b) Arrangement of petal in endcaps ~\cite{TDR}}
\label{fig:stavelayout}
\end{figure}

\section{Modules}
The basic unit of the ITk Strip Detector is the silicon-strip module. As shown in Fig.~\ref{fig:module}, a module consists of one sensor and one or two low-mass PCB’s, called hybrids, hosting the read-out ASICs (ABCStar and HCCStar) and one power board.  Due to the large number of modules required for the ITk Strip Detector, the modules have been designed with mass production and low cost in mind.

The ITk Strip modules are constructed by directly gluing kapton flex hybrids and power boards to silicon sensors with electronics-grade epoxy. Readout ASICs are then wire bonded to the sensor. Various strip lengths and geometries are foreseen, depending on the planned location of the module within the detector.

A power poard consists of a DC-DC Low Voltage (LV) power block, monitoring ASICs, and a high-voltage multiplexer~\cite{HVMX}. A prototype of power board is shown in Fig.~\ref{fig:module}.  
A custom ASIC called the Autonomous Monitor Chip (AMAC) will monitor powering and environmental conditions. Based on the measured conditions, the AMAC will control both the high and low voltage power of each module independently.
\begin{figure}[]

  \centering
    \subfigure[]{\includegraphics[width=0.49\textwidth]{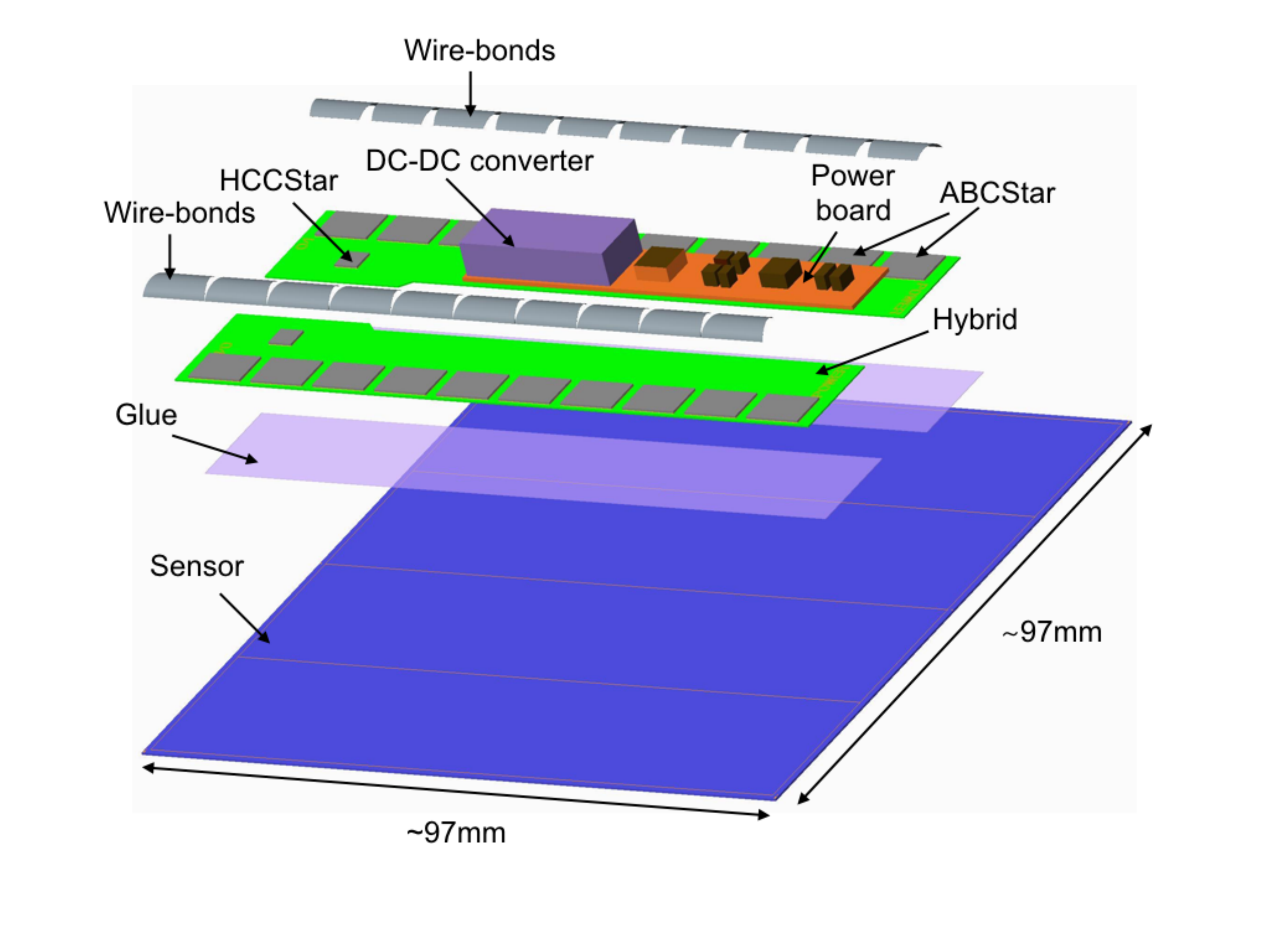}}
    \subfigure[]{\includegraphics[width=0.49\textwidth]{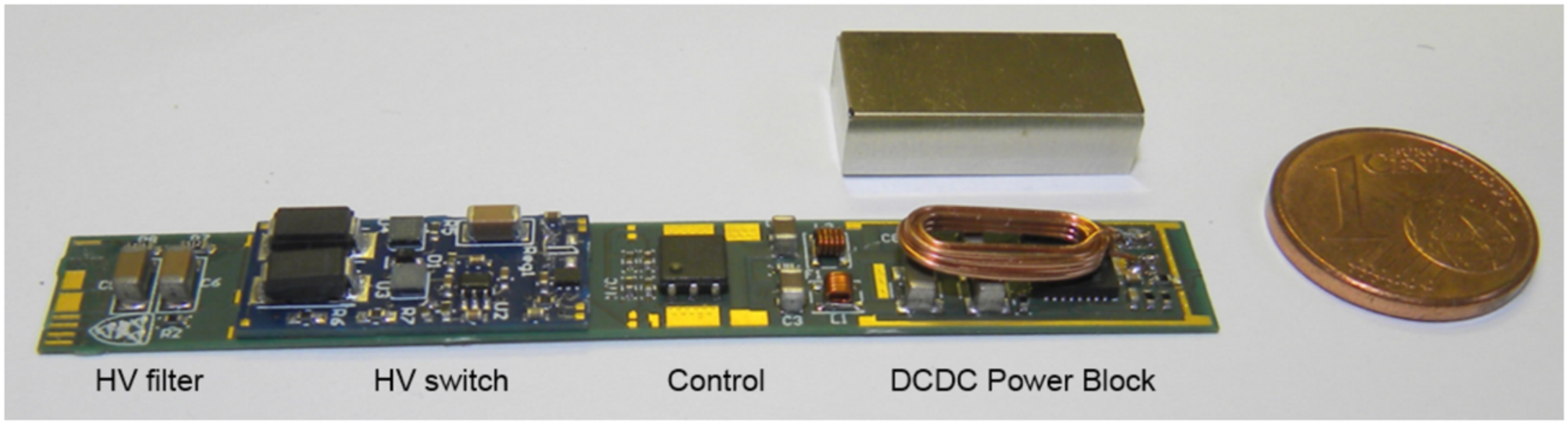}}
  \caption{(a)Exploded view of a short-strip barrel module with all relevant components~\cite{TDR}. (b) A prototype of power board. }
\label{fig:module}
\end{figure}

\section{Sensors}
The current ATLAS strip detector uses silicon strip sensors made with p-strips implanted on n-type silicon bulk (p-in-n), and it is designed to operate up to 2$\times 10^{14} n_{eq}$/\cm$^2$. In order to maintain the detector performance for the full HL-LHC lifetime, new sensors are needed to operate when exposed to particle fluences of up to 2$\times 10^{15} n_{eq}$/\cm$^2$.

The baseline proposal for the new radiation hard sensor is called n$^+$-in-p type sensor~\cite{ninp}. 
In this sensor, the strips are AC-coupled with n-type implants in a p-type float-zone silicon bulk (n$^+$-in-p FZ) as shown in Fig.~\ref{fig:sensor} (a).
This type of sensor collects electrons and has no radiation induced type inversion.
There are two types of barrel sensors: one has four rows of short strips in the $z$-direction (24.10 $\mathrm{mm}$) and the other has two rows of longer strips (48.20 $\mathrm{mm}$) for the short-strip and long-strip modules, respectively.
The petal sensors require radial strips (i.e. pointing to the beam-axis) to give a measurement of the $R\phi$ coordinate. As a result, these sensors have a wedge shape with curved edges.

The radiation hardness of these prototype sensors has been measured. The charge collected in prototype sensor ATLAS12~\cite{ATLAS12} at 500 V bias voltage as a function of the fluence after irradiation with different particles is shown in Fig.~\ref{fig:sensor} (b). Tests indicate a total drop in collected charge of a factor of ∼2 at end-of-life doses~\cite{TDR}. The radiation hardness of the sensor is good enough to maintain the strip detector performance at the end of HL-LHC operation.

\begin{figure}[]

  \centering
    \subfigure[]{\includegraphics[width=0.49\textwidth]{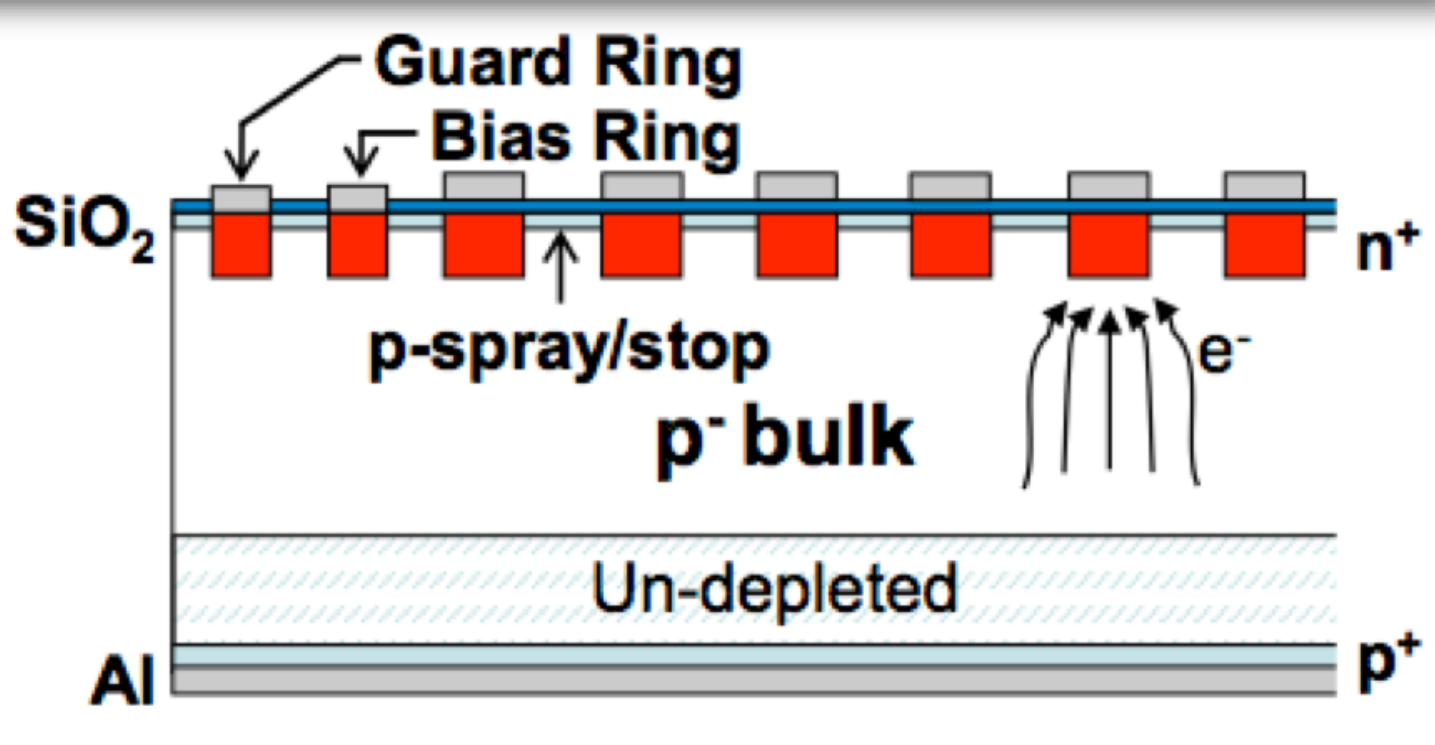}}
    \subfigure[]{\includegraphics[width=0.49\textwidth]{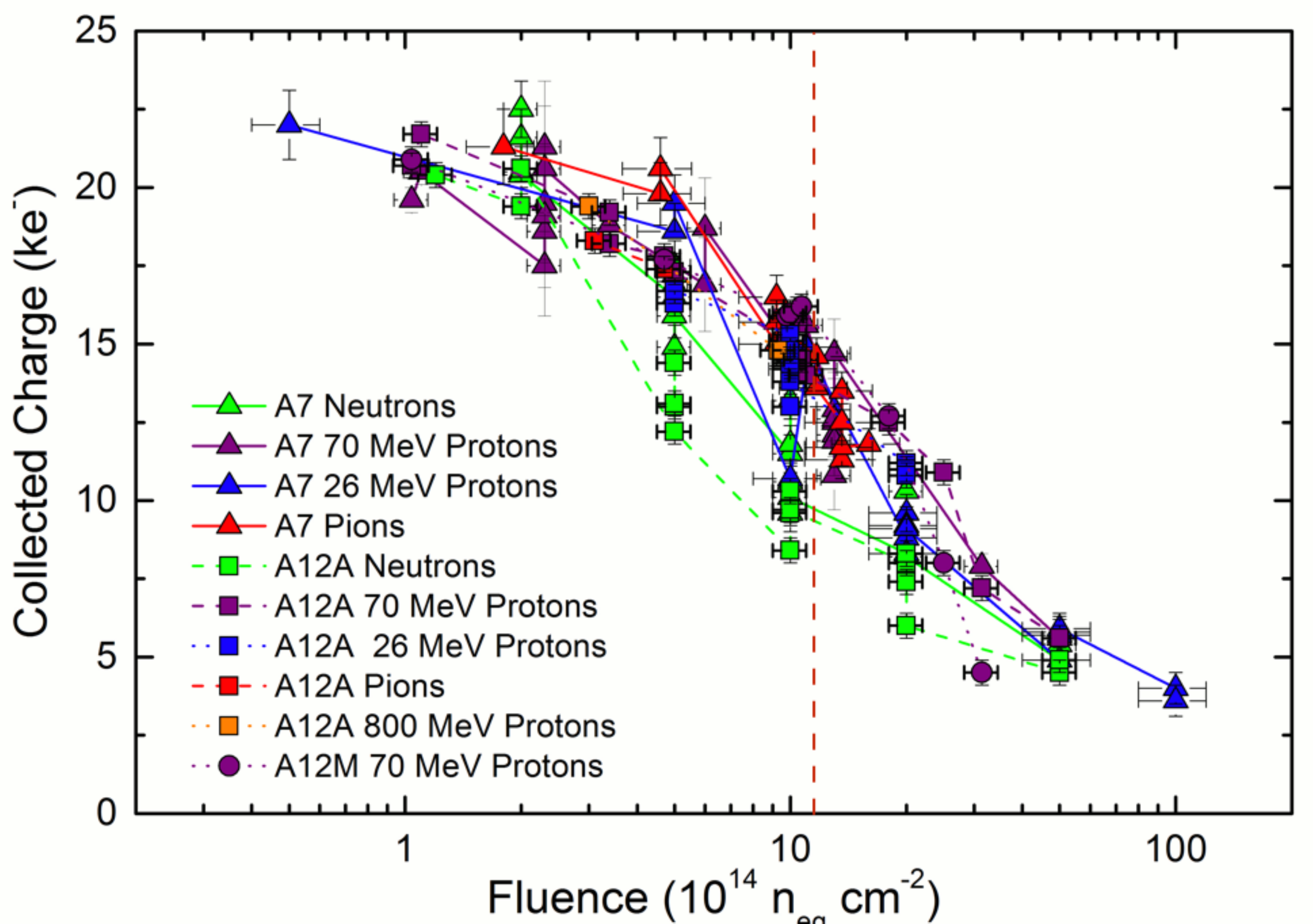}}
  \caption{(a) Cross section of (n$^+$-in-p) strip sensors in baseline design (b) Collected signal charge in prototype sensor ATLAS12 at 500 V bias voltage for minimum ionising particles as a function of 1~\MeV~$n_{eq}$/\cm$^2$ fluence for various types of particles~\cite{sensorTID}. The red dash curve stands for the expected maximum fluence in strip detector volume at the end of HL-LHC lifetime.   }

\label{fig:sensor}
\end{figure}

\section{Overall Electronics Architecture}

\begin{figure}[]

  \centering
    \subfigure[]{\includegraphics[width=0.8\textwidth]{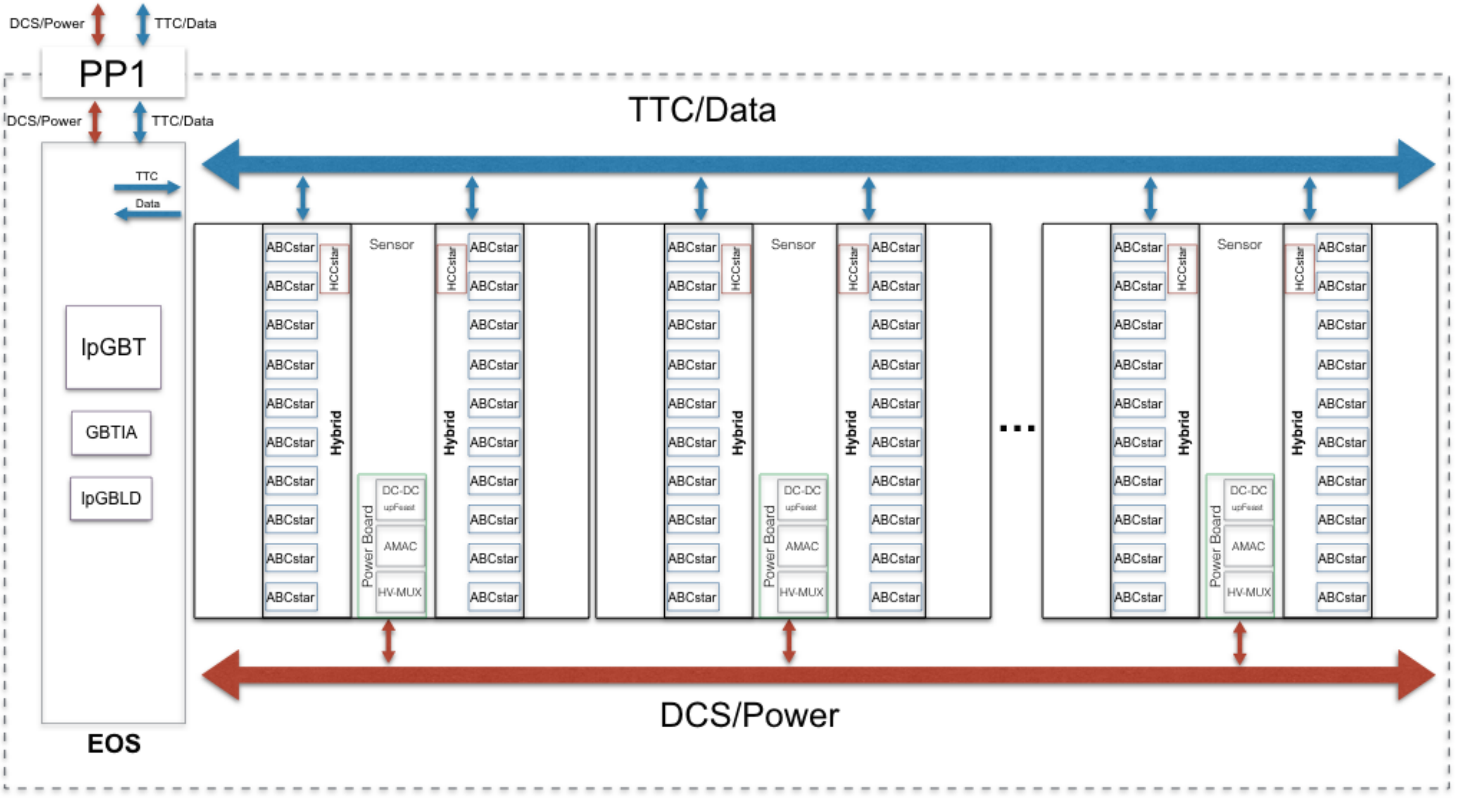}}
    \subfigure[]{\includegraphics[width=0.8\textwidth]{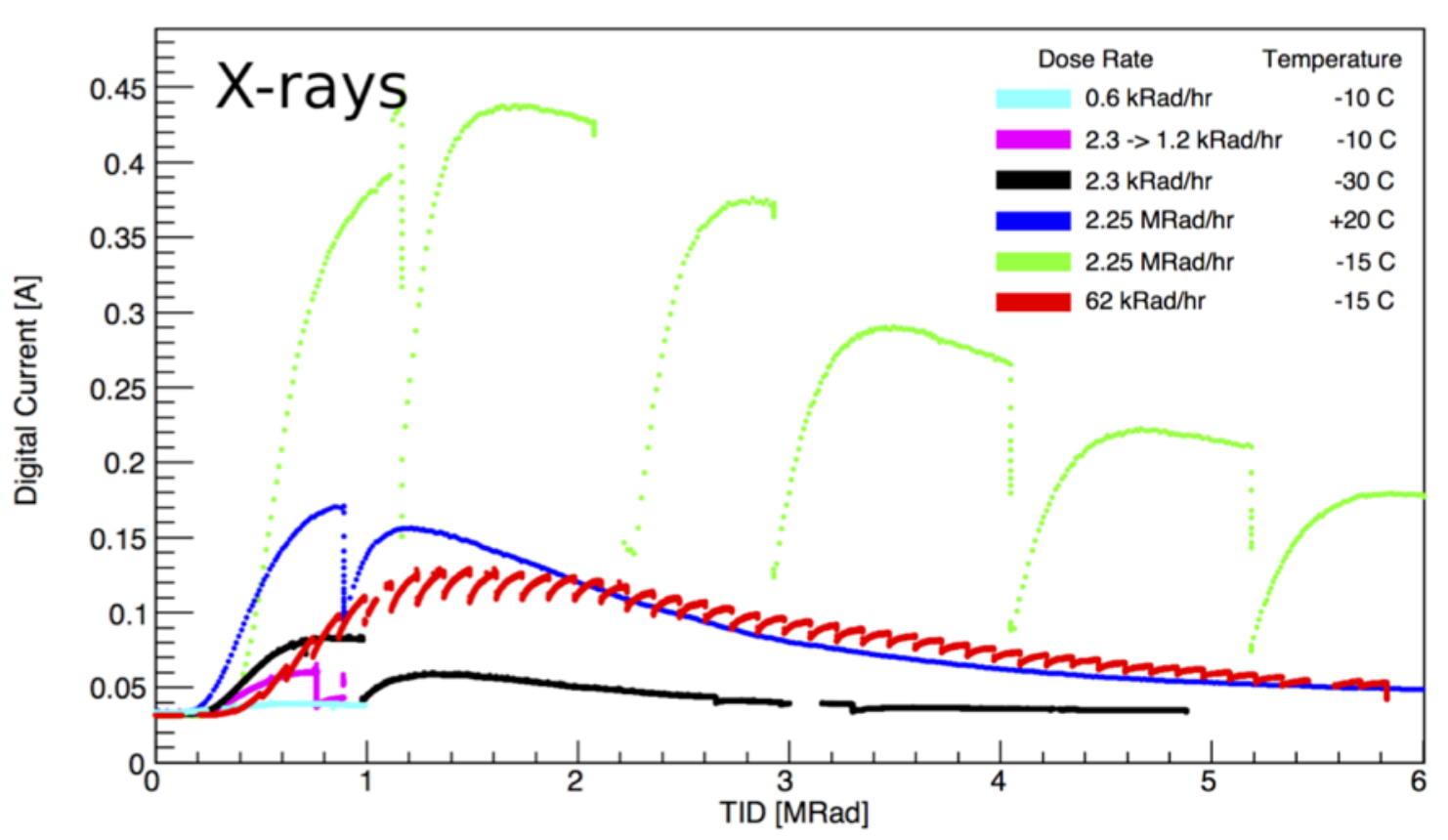}}

  \caption{ (a) Schematic of the power and electronics architecture. The Timing, Trigger, and Control (TTC), Detector Control Signals (DCS), and power are carried between the modules and the End of Substructure (Eos) card via custom bus tapes. Optical links are used for communication lines between the EoS and the off-detector readout. (b)Digital current vs. TID for ABC130 chips during X-rays irradiations at different dose rates and temperatures~\cite{TDR}.}
\label{fig:readout}
\end{figure}

Charged particles passing through the sensor create a signal charge within the silicon sensor diode. As shown in Fig.~\ref{fig:readout} (a), this signal is transmitted through a wire to the front-end chip, the ATLAS Binary Chip (ABCStar) containing 256 pre-amplifiers and discriminators together with the Level 0 buffer, event builder and cluster finder. With the front-end chip ABCStar the signal on each channel is amplified, shaped and then discriminated to provide a binary output. Depending on the sensor type, up to 12 ABCStar ASICs are grouped on to one hybrid. Each hybrid has a Hybrid Controller Chip (HCCStar) that interfaces the stave/petal service bus and the front-end ASICs on the Strip Detector hybrids. The HCCStar receives the signals from the ABCStar, builds packets and then transmits them onto the EOS. It also receives the clock and control signals (TTC) and distributes those to the ABCStar.

Trigger, Timing, and Control (TTC) signals arriving from the off-detector systems are sent from the EoS to each HCCStar via the TTC bus on the bus-tape. The EoS includes a low power GigaBit Transceiver (lpGBTx~\cite{GBT}) that interfaces with the HCC130 ASICs and a Versatile link (VTRx+~\cite{versatile}) fibre optic driver.

The design of the readout chip is ongoing. The current version of the chip is the ABC130. The final design of the chip, called the ABCStar, is expected to be submitted in 2018. The ABC130 has undergone tests to ensure its ability to deal with both the high flux rates and the large dose it will have accumulated by the end of HL-LHC lifetime. 

As illustrated by Fig.~\ref{fig:readout} (b), the tests showed a large increase in the current drawn by the digital portion of the chip around a Total Ionising Dose (TID) of 1 MRad. The current then slowly returns to its nominal value with increasing dose. These tests also confirmed the increase in current is correlated with dose rate and anti-correlated with temperature as predicted by an empirical model. Tests have been performed at dose rates and temperatures compatible with HL-LHC expectations and show an increase of order 50\% -- 100\% in the digital current for the chips in areas of the detector which will receive the highest dose rate. Considerations for cooling and powering requirements have been adjusted to take into account the increased power consumption of the chips at the TID bump.





\section{Beam Tests}
A series of tests have been performed at the DESY-II and CERN SPS test beam facilities to investigate the detailed performance of the strip module before and after irradiation. The DESY tests used electrons with an energy of 4 to 4.8 GeV, and the setup of the DESY test beams is shown in Fig.~\ref{fig:teambeam} (a).  non-irradiated modules and irradiated modules operated successfully in test beams.

The efficiency curves for long-strip regions of seven ASICs of the non-irradiated module and one ASIC on the irradiated module are shown in Fig.~\ref{fig:teambeam} (b)~\cite{TDR}.
The signal-to-noise ratio of an irradiated strip module is measured to be larger than 20.
According to the test beams results, modules are expected to have greater than 99\% detection efficiency at thresholds that allow for operation with less than $1 \times 10 ^{-3}$ channel noise occupancy at the end of HL-LHC lifetime. 

\begin{figure}[]

  \centering
    \subfigure[]{\includegraphics[width=0.49\textwidth]{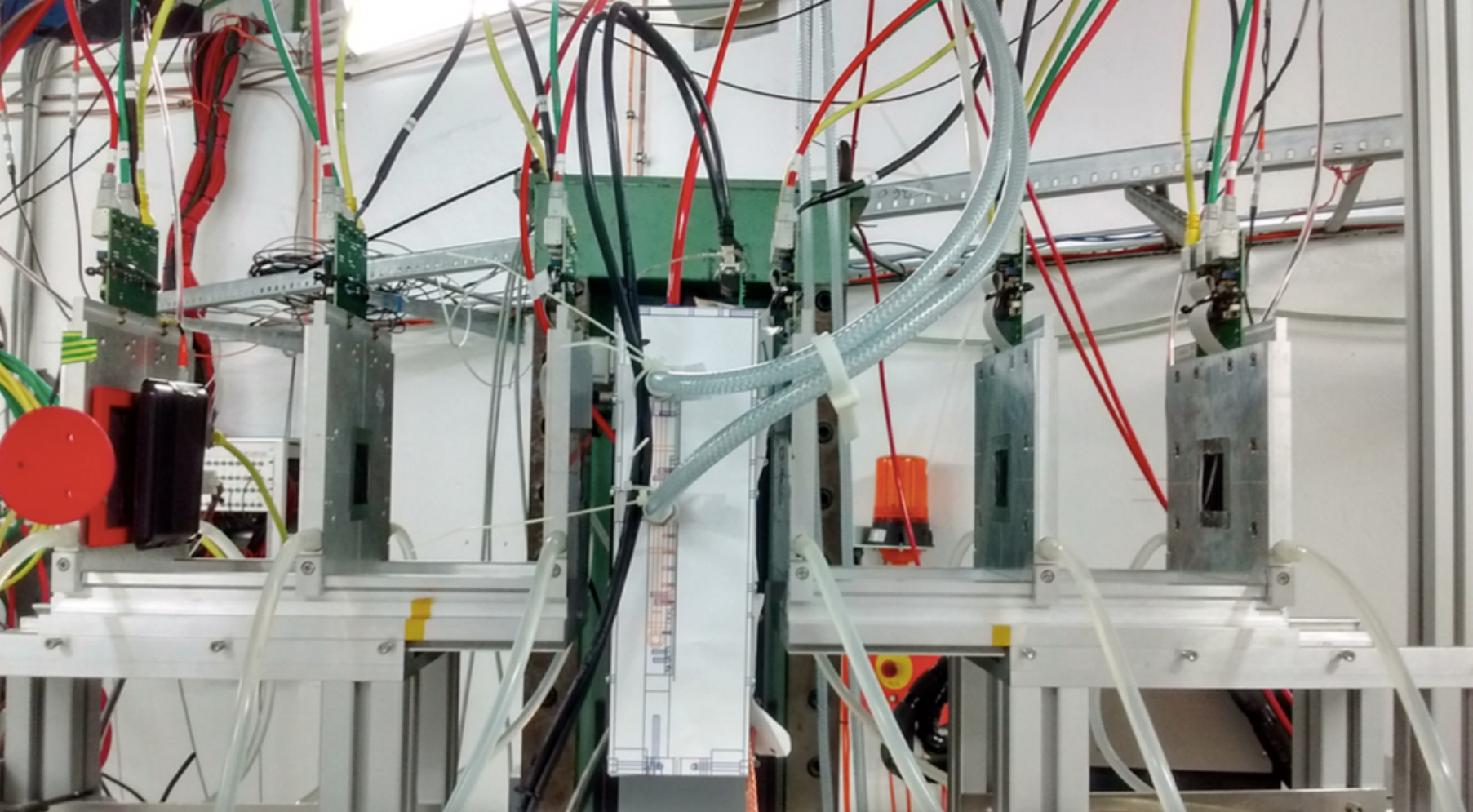}}
    \subfigure[]{\includegraphics[width=0.49\textwidth]{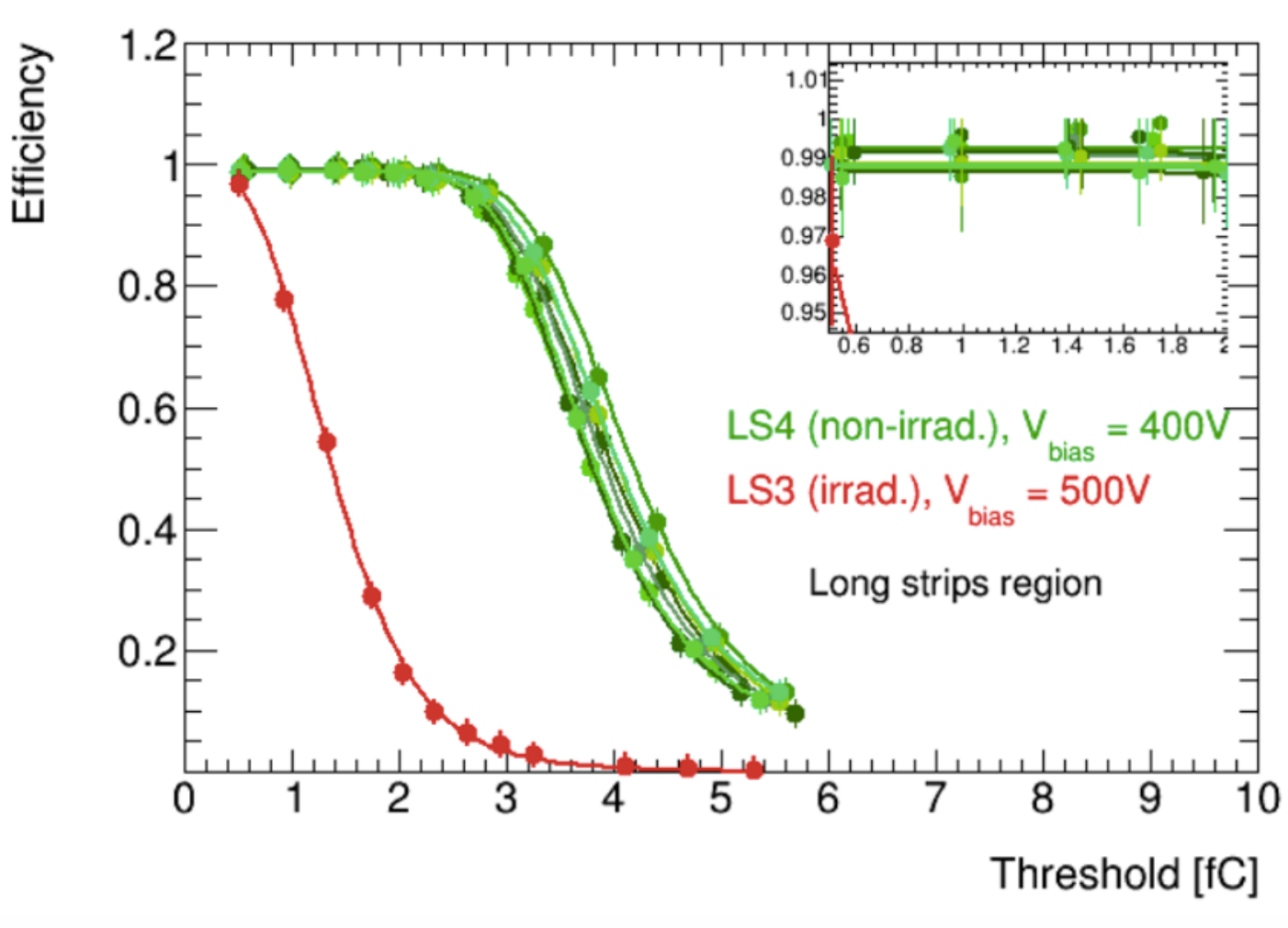}}
  \caption{(a) The DURANTA Telescope at DESY with a barrel mini module under test. (b) The efficiency versus the threshold for four ASICs on the non-irradiated long-strip module (shown in green), and one ASIC on the irradiated module (shown in red).}

\label{fig:teambeam}
\end{figure}

\section{Summary }
 A new all-silicon tracker will replace the current ATLAS tracking detector for HL-LHC data taking, as driven by the tracking performance requirements. The layout, expected performance and latest status of prototyping of a new silicon strip tracker for the ATLAS experiment at the HL-LHC has been shown.  
A modular design approach where small, largely independent building blocks form large structures has been employed. Testing of relevant components is underway and challenges are being addressed.

\section{Acknowledgements}
The author was supported and financed by Chinese National Key Program for S \& T Research and Development(Grant No. 2016YFA0400101).

\end{document}
\endinput